\documentclass{aa}
\usepackage{graphics}
\begin{document}
\title{The new sample of giant radio sources\\
III. Statistical trends and correlations}
\author{J. Machalski, and M. Jamrozy}
\offprints{J. Machalski}
\institute{Astronomical Observatory, Jagellonian University,
ul. Orla 171, PL-30244 Cracow, Poland\\
email: machalsk@oa.uj.edu.pl}
\date{Received ....; accepted ...}
\abstract{}
{In this paper we analyse whether `giant' radio galaxies (GRGs) differ from
`normal'-size galaxies (NSGs) except for the linear extent of their radio structure.}
{We compare a number of properties of GRGs with the corresponding properties of
NSGs, and analyse the statistical trends and correlations of physical parameters,
homogeneously determined for the sources, with their `fundamental' parameters:
the redshift, radio luminosity, and linear size. Using the Pearson
partial-correlation test on the correlation between two variables in the presence
of one or two other variables, we examine which correlation is the strongest.}
{The analysis clearly shows that GRGs do not form a separate class of radio sources.
They most likely evolve with time from smaller sources, however under specific
circumstances. Analysing properties of GRGs and NSGs together, we
find that (i) the core prominence does not correlate with the total radio
luminosity (as does the core power), but it anti-correlates
with the surface brightness of the lobes of sources, (ii) the energy density
(and possibly the internal pressure) in the lobes is independent of redshift for
constant radio luminosity and size of the sources. Thus, in the analysed samples,
there is no evidence for a cosmological evolution of the IGM pressure in the form
$p_{\rm IGM}\propto (1+z)^{5}$, (iii) the equipartition magnetic-field strength,
transformed into constant source luminosity and redshift, strongly correlates
with the source size. We argue that this $B_{\rm eq}$--$D$ correlation reflects
a more fundamental correlation between $B_{\rm eq}$ and the source age, (iv) both
the rotation and depolarisation measures suggest Faraday screens local to the lobes
of sources, however their geometry and the composition of intervening material
cannot be determined from the global polarisation characteristics. The significant
correlation between the depolarisation measure and the linear size can be explained
by less dense IGM surrounding the lobes (or cocoon) of GRGs than that in the
vicinity of NSGs.}
{}
\authorrunning{J. Machalski \& Jamrozy}
\titlerunning{Giant radio sources, trends and correlations}
\maketitle

\section{Introduction}

Classical double radio sources with projected linear size greater than 1 Mpc are
commonly referred to as `giants'; this size limit was based on the cosmological
constants  $H_{0}=50$\,km\,s$^{-1}$Mpc$^{-1}$ and $q_{0}=0.5$. After adopting a flat
Universe with $H_{0}=71$\,km\,s$^{-1}$Mpc$^{-1}$ and $\Omega_{m}=0.27$, the above size
limit is reduced to about 700 kpc. For consistency with many previously published
papers, in the present analysis we include FRII-type sources with $D\geq 700$ kpc
into the sample of giants.

In Paper I (Machalski et al., 2001) we selected a sample of 36 giant radio source
candidates, primarily of FRII-type morphology (Fanaroff \& Riley 1974), and we
presented their optical identifications and low-resolution spectra used to determine
the object's redshifts. This in turn allowed us to derive a number of physical
parameters for the sample sources, like projected linear size, radio luminosity,
optical absolute magnitude of identified host galaxy, equipartition energy density
and magnetic field strength, etc. In Paper II (Machalski et al., this volume) the
previously published data were supplemented with high-frequency total-intensity
and polarised-intensity radio maps, and the polarisation and depolarisation parameters
of the sample sources  were specified.

In this paper we compare these physical parameters determined for an enlarged
sample of giant radio galaxies with the corresponding parameters in a comparison
sample of normal-size FRII-type radio galaxies, i.e. samples which do not
comprise quasars with extended double radio structures. Here we
analyse properties of the whole radio sources. A further analysis of the
sample sources' asymmetries, properties of their lobes, etc., will be given in a
forthcoming paper.
The observational data used is described in Sect.~2. Statistical trends and
correlations between different parameters of the sources are analysed in Sect.~3,
while the results are discussed and summarised  in Sect.~4.

\section{The data}

\subsection{Giant-sized radio galaxies}

The sample consists of 28 giant-sized galaxies out of the 36 the sources presented
in Paper II, and 15 of 18 giant radio sources selected from the paper of Machalski,
Chy\.{z}y \& Jamrozy (2004).
The redshift range of the sample sources is  $0.06<z<0.82$ with a median
value of 0.26$\pm$0.03, and mean deviations from the median of $-$0.10,+0.21 
(concerning an asymmetrical distribution).
The 1.4-GHz luminosity $P_{1.4}$[W\,Hz$^{-1}]$ has log values in the range 
$24.3<\log P_{1.4}<27.3$ with a median of 25.6$\pm$0.07 and  mean deviations of
$-$0.43,+0.47.
For all these 43 sources their geometry, radio spectrum, lobe brightness, arm
ratio, core prominence, and equipartition energy density, internal
pressure and magnetic field strength are homogeneously determined. 17 of the
above 43 galaxies form the giant subsample, for which polarisation and
depolarisation parameters are available from Paper II. For simplicity, giant
radio galaxies are referred to hereafter as GRGs.

\subsection{Normal-sized galaxies}

The comparison sample consists of 75 FRII-type sources for which the published
data allowed a determination of the same parameters as for the sample of giant
radio galaxies. The sources are selected to fulfill the following criteria:

\noindent
-- have the radio core detected, 

\noindent
-- asymmetries in their arm-ratio, and luminosity and spectral index of the lobes,
can be determined from the published maps,

\noindent
-- polarisation data are available in some cases.

As a result, the sample comprises (i) high-luminosity low-redshift 3CR sources
selected from the papers of Leahy \& Perley (1991), and Hardcastle et al. (1998),
(ii) high-luminosity high-redshift 3CR, as well as low-luminosity low-redshift B2
sources used by Machalski et al. (2004) as a comparison sample for their giant
radio galaxies sample. Since the desired polarisation data are limited to a fraction
of these sources only, we include also (iii) southern radio galaxies selected from
the Molonglo survey by Ishwara-Chandra et al. (1998). The latter galaxies are
chosen mostly for their polarisation and depolarisation data given in that paper. 
The redshift range in our comparison sample is $0.03<z<1.8$ with a median value of
0.26$\pm$0.05, and the mean deviations from the median of $-$0.14,+0.55. The 1.4-GHz
(log) luminosity range is  24.3$<\log P_{1.4}$[W\,Hz$^{-1}]<$28.6. A median value
of the distribution is 26.8$\pm$0.02, and mean deviations of $-$1.0,+1.0. For 47
of the 75 sources the polarisation and depolarisation parameters, similar to those in
the GRG sample, were available from Garrington et al. (1991), Ishwara-Chandra et al.
(1998), and Goodlet et al. (2004). Hereafter normal-sized radio galaxies are
referred to as NSGs.

\section{The analysis and results}
\subsection{The method}

The aim of our analysis is to investigate any trends and/or correlations between
physical parameters determined for the sample sources and the `fundamental'
parameters: the redshift, $z$, radio luminosity at 1.4 GHz, $P_{1.4}$, and the
linear size, $D$.
The method applied in the present study is based on the homogeneous determination
of a number of observational and physical parameters ($m$) for all ($N$) members of
both samples, and then inserting these into a numerical array of $m\times N$
elements. Most of these parameters are interdependent, hence each parameter of the
sample sources correlates somehow with the other parameters. Therefore, given the
array, a statistical test for correlations between two variables in the presence
of one or two other variables is used to examine relations among the properties of
giant and normal-sized radio galaxies. In order to determine which correlation
is the strongest, whether a third (or a third and a fourth) variable causes the
correlation between the other two, and whether there is a residual correlation
between these two variables when the third (or the third and fourth) is (are) held
constant, we calculate the Pearson partial correlation coefficients for the
correlation between the selected parameters.

Due to the fact that many correlations between different parameters seem to
follow a power law, two numerical arrays are used: the first with primary values
of some of these parameters, and  the other with their logarithms. For example:
$D$[kpc] and $1+z$ are in the first array; log($D$[kpc]) and log$(1+z)$ are in the
other. Hereafter $r_{XY}$ denotes the correlation coefficient for the correlation
between parameters $X$ and $Y$ (hereafter referred to as the `direct' correlation),
$r_{XY/U}$ is the partial correlation coefficient between these parameters in
the presence of a third parameter, $U$, which can correlate with both $X$ and $Y$, and
${\cal P}_{XY/U}$ is the probability that the test pair $X$ and $Y$ is uncorrelated
when $U$ is held constant. Similarly, $r_{XY/UV}$, ${\cal P}_{XY/UV}$ is the
correlation coefficient for a correlation involving four parameters, and the related
probability, respectively.

\subsection{Radio core properties}

In this subsection, we analyse the Pearson partial correlations between each of two
radio core parameters: the core power and the core prominence, and other
parameters of the sample sources which give the highest $r_{XY}$. A useful
measure of the core prominence is the ratio $c_{\rm p}$=
$S_{\rm core}/(S_{\rm total}-S_{\rm core})$ (cf. Morganti et al., 1993).
Similarly to Lara et al. (2004), we derive this parameter using $S_{\rm core}$
measured at 5 GHz and $S_{\rm total}$ at 1.4 GHz.

\subsubsection{The core power partial correlations} 

The strong correlation between the core power at 5 GHz and the total power at lower
frequencies in the population of classical double radio sources is very well known
(cf. Giovannini et al., 2001). This correlation can be attributed to the Doppler
beaming of a parsec-scale jet (e.g. Bicknell 1994; Komissarov 1994) not discerned
from the core with a medium (VLA) angular resolution, and can reflect different
inclination angle of the nuclear jets, and thus the inclination of the entire radio
source's axis to the observer's line of sight. In this case, relatively stronger
cores should be observed in more strongly projected sources. Therefore, in giant
radio galaxies, with the inclination angle very likely close to 90$\degr$, one could
expect to observe relatively weaker cores which is not the case (cf. Lara et al., 2004).

Our data support the previous results, and we find that the direct correlation
coefficient between log$P_{5}^{\rm core}$ and log$P_{1.4}$ is high. Nevertheless,
the core power in our samples also  correlates with other physical parameters; in
order of decreasing $r_{XY}$: the redshift and linear size of the source, $D$.
The partial correlation coefficients in the correlation of log$P_{5}^{\rm core}$
with log$P_{1.4}$, log($1+z$), and log$D$ together with the related probabilities
of their chance correlation are given in Table~1.

\begin{figure}[t]
\resizebox{\hsize}{!}{\includegraphics{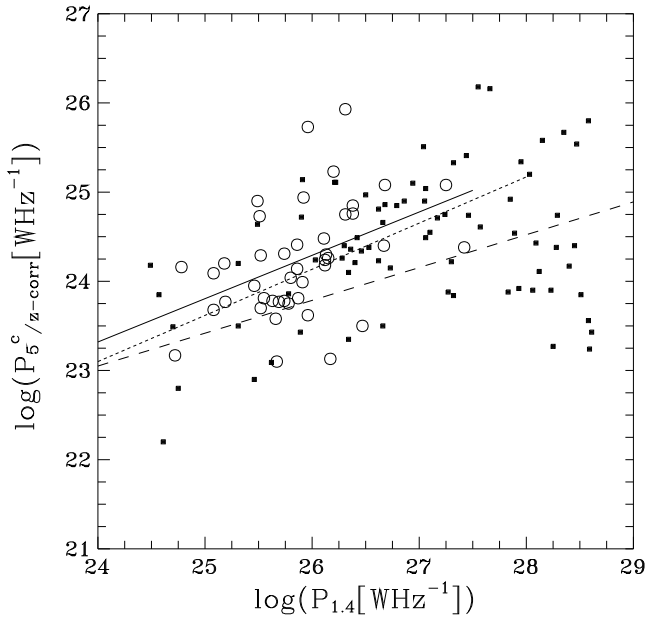}}
\caption{Core power at 5 GHz transformed to the reference redshift of 0.5 vs. total
power at 1.4 GHz. GRGs are marked
with open circles and NSGs with small squares. The solid line indicates a least
squares fit to the GRGs data. The dashed line shows the fit to the NSGs data, and
the dotted line -- the fit to the NSGs with the same luminosity range as the GRGs}
\end{figure}

\begin{table}[h]
\caption{The correlation of core (log) luminosity $P_{5}^{\rm c}$ with 
$P_{1.4}$, or 1+$z$, or $u_{\rm eq}$ when other parameters are held constant}
\begin{tabular}{lllll}
\hline
Correlation & $r_{XY}$ & $r_{XY/U}$ & ${\cal P}_{XY/U}$ \\
N=118     & & $r_{XY/V}$ & ${\cal P}_{XY/V}$ & $r_{XY/UV}$\\
& & & &   ${\cal P}_{XY/UV}$\\
\hline 
$P_{5}^{\rm c}-P_{1.4}/D$ & +0.703 & +0.711 & $\ll$0.0001\\
$P_{5}^{\rm c}-P_{1.4}/1+z$      & & +0.537 & $\ll$0.0001\\
$P_{5}^{\rm c}-P_{1.4}/D$,1+z & & & & +0.570\\
 & & & &  $\ll$0.0001\\
$P_{5}^{\rm c}-$(1+z)/$P_{1.4}$ & +0.538 & +0.021 & 0.82\\
$P_{5}^{\rm c}-$(1+z)/$D$ & & +0.518 & $\ll$0.0001\\
$P_{5}^{\rm c}-$(1+z)/$P_{1.4},D$ & & & & +0.006\\
 & & & &  0.95\\
$P_{5}^{\rm c}-D/P_{1.4}$ & $-$0.172 & +0.226 & 0.015\\
$P_{5}^{\rm c}-D/$1+z   & & $-$0.016 & 0.86\\
$P_{5}^{\rm c}-D/P_{1.4},$1+z & & & & +0.225\\
 & & & &  0.015\\
\hline
\end{tabular}
\end{table}

The above tests confirm the strong log$P_{5}^{\rm core}$--log$P_{1.4}$
correlation, and completely exclude any significant dependence of the core
power on redshift, when $P_{1.4}$ and $D$ are held constant.
Fitting a surface to the values of log$P_{5}^{\rm core}$ over the
log$P_{1.4}$--log($1+z$) plane (where $P_{1.4}$ is in W\,Hz$^{-1}$), we found

\begin{equation}
P_{5}^{\rm core}\propto P_{1.4}^{0.55\pm0.08}(1+z)^{0.29\pm0.08}.
\end{equation}

\noindent
Note that the power of 0.55 is lower than that in the Giovannini et al.'s relation
transformed to the cosmological constants adopted in this paper, $P_{5}^{\rm core}
\propto P_{\rm t}^{0.60\pm0.04}$ (cf. Paper II), and Giovannini et al. do not take
into account the dependence of the total power $P_{\rm t}$ on redshift.

Using Eq.\,(1) we eliminate dependence of the core power on redshift
transforming its values to a reference value of $z$. The plot of log$P_{5}^{\rm core}$
transformed to $z$=0.5 as a function of log$P_{1.4}$ is shown in Fig.~1. The sample
GRGs are indicated by open circles, and the NSGs  by small full squares. The solid
and dashed lines show  formal linear regressions of log$P_{5}^{\rm core}$ on the
log$P_{1.4}$ axis for GRGs and NSGs, respectively. Although these regression lines
suggest a trend of the GRGs cores to be more powerful as compared with the NSG cores
of the same total radio power $P_{1.4}$, statistical tests indicate that differences
between both the slopes and the $P_{5}^{\rm core}$ intercepts are statistically
insignificant. The probability of being drawn from the same general population is
between 40\% and 60\%. The difference between these regression lines almost
disappears when NSG and GRG galaxies within the same total power range are compared,
as indicated by the dotted line for the NSGs with $P_{1.4}<10^{27.3}$ W\,Hz$^{-1}$.

\subsubsection{The core prominence partial correlations}

The correlation coefficients in the correlations of $c_{\rm p}$ with the
source fundamental parameters indicate strong (by definition) anticorrelation
with the source power $P_{1.4}$, and strong correlation with its size $D$.
However, we find that the core prominence most strongly (anti)correlates with
the source surface brightness, defined here as ${\cal B}=P_{1.4}/(D^{2}/AR)$,
where $AR$ is the source (its cocoon) axial ratio (for its definition cf. Paper II),
and the source size is given in metres.

The partial correlation coefficients in the correlation of log\,$c_{\rm p}$ with
log\,${\cal B}$, log$P_{1.4}$, and log(1+$z$) together with the corresponding
probabilities of their chance correlation are given in Table~2.

\begin{figure}[t]
\resizebox{\hsize}{!}{\includegraphics{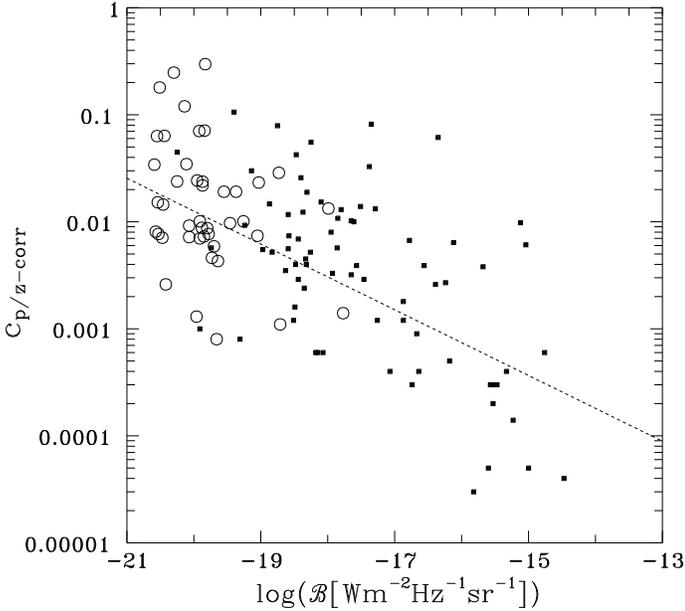}}
\caption{Core prominence transformed to the reference redshift of 0.5 vs. source
surface brightness. GRGs and NSGs are marked with the same symbols as in Fig.\,1.
The dashed line shows the least squares fit to the entire data}
\end{figure}

\begin{table}[h]
\caption{The correlation of core (log) prominence $c_{\rm p}$ with ${\cal B}$,
or $P_{1.4}$, or 1+$z$,  when other parameters are held constant}
\begin{tabular}{lllll}
\hline
Correlation & $r_{XY}$ & $r_{XY/U}$ & ${\cal P}_{XY/U}$ \\
N=118     & & $r_{XY/V}$ & ${\cal P}_{XY/V}$ & $r_{XY/UV}$\\
& & & &   ${\cal P}_{XY/UV}$\\
\hline
$c_{\rm p}-{\cal B}/P_{1.4}$ & $-$0.643 & $-$0.315 & $<$0.001\\
$c_{\rm p}-{\cal B}/$1+$z$ & & $-$0.548 & $\ll$0.0001\\
$c_{\rm p}-{\cal B}/P_{1.4}$,1+$z$ & & & & $-$0.313\\
 & & & & $<$0.001\\
$c_{\rm p}-P_{1.4}/{\cal B}$ & $-$0.560 & $-$0.034 & 0.72\\
$c_{\rm p}-P_{1.4}/$1+$z$    & & $-$0.419 & $<$0.0001\\
$c_{\rm p}-P_{1.4}/{\cal B}$,1+$z$ & & & & $-$0.043\\
& & & & 0.64\\
$c_{\rm p}-$(1+$z$)/${\cal B}$ & $-$0.410 & +0.040 & 0.67\\
$c_{\rm p}-$(1+$z$)/$P_{1.4}$  & & +0.019 & 0.84\\
$c_{\rm p}-$(1+$z$)/${\cal B},P_{1.4}$ & & & & +0.029\\
& & & & 0.76\\
\hline
\end{tabular}
\end{table}

Whereas the core prominence most strongly correlates with the surface brightness,
the partial correlation coefficients in Table~2 show that its dependences on the source's
total power as well as on redshift are marginal when the surface brightness is kept
constant. Fitting a surface to the values of $c_{\rm p}$ over the
log\,$P_{1.4}$--log($1+z$) plane, we find

\begin{equation}
c_{\rm p}\propto P_{1.4}^{-0.23\pm 0.05}(1+z)^{+3.00\pm 0.35}.
\end{equation}

\noindent
The values of log\,$c_{\rm p}$, transformed to $z$=0.5 vs. log\,${\cal B}$ is presented
in Fig.~2.

\subsubsection{The core prominence and the orientation indicator}

In the sample of Lara et al. (2004) comprising of large size FRI and FRII-type sources, the
authors found an excess of sources with a core power larger than expected from their total 
power, and considered whether the ratio of $P_{5}^{\rm core}$ and $P_{\rm norm}^{\rm core}$,
i.e. that calculated from the relation of Giovannini et al. (cf. Sect.\,3.2.1), might be
an indicator of the source orientation angle (${\cal P}_{CN}$ in their paper).
For sources larger than 1 Mpc they found the median ratio of ${\cal P}_{CN}\approx 1.6$. 

Our sample confirms the above effect, though qualitatively only. Using their
normalization transformed to the cosmological constants adopted in our samples, the 
median of ${\cal P}_{CN}$ for GRGs and NSGs is
$0.89^{+0.21}_{-0.13}$ and $0.29^{+0.04}_{-0.03}$, respectively. The distributions of
log\,${\cal P}_{CN}$ for the sample GRGs and NSGs are shown in Fig.~3.  A value of
${\cal P}_{CN}<1$ for NSGs is justified because our comparison sample of radio sources
does not include quasars, for which the core power is statistically higher than that
for radio galaxies. On the other hand, the median of ${\cal P}_{CN}$ close to unity
supports the result described in Sect.\,3.2.1 that radio cores of giant-size radio
galaxies are not statistically stronger than those for normal-size galaxies.

\begin{figure}[t]
\resizebox{\hsize}{!}{\includegraphics{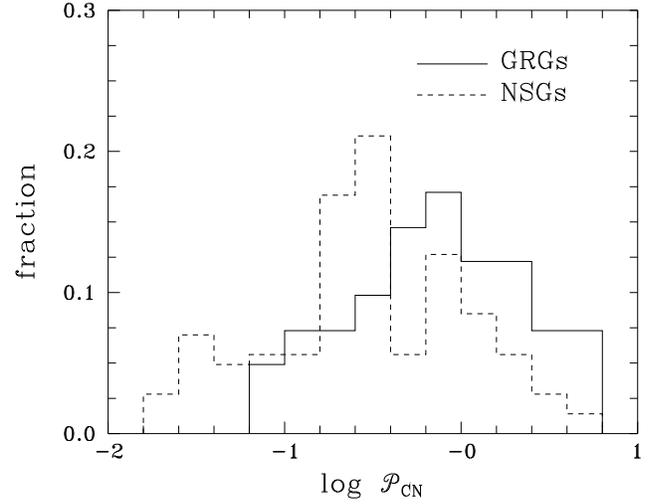}}
\caption{Distribution of log\,${\cal P}_{CN}$ for the Giant (GRG: solid line) and
`normal-size' (NSG: dashed line) radio galaxies}
\end{figure}

\subsection{Equipartition energy density and magnetic field strength}

Two other physical parameters of the sample sources derived directly from the
observational data are: the equipartition energy density, $u_{\rm eq}$, and magnetic
field strength, $B_{\rm eq}$. The values of these two parameters for the sample
sources have been calculated using the formulae of Miley (1980), and assuming the
ratio of energy in protons to that in electrons $k$=1, and the filling factor
$\eta$=1 (cf. Paper II).
Formally, we analyse relations of $u_{\rm eq}$ and $B_{\rm eq}$  with the sources'
radio luminosity, size, and redshift.  However, the equipartition energy density
and corresponding magnetic field are related, by definition, to the luminosity and
size with the canonical formulae $u_{\rm eq}\propto P^{4/7}V^{-4/7}$ (i.e.
$u_{\rm eq}\propto P^{4/7}D^{-12/7}$), and $B_{\rm eq}\propto u_{\rm eq}^{1/2}$,
respectively.

\subsubsection{Energy density partial correlations}

Our statistical analysis, involving the largest sources known, shows that besides
the expected strong  correlation between the energy density and the luminosity
of sources, and anti-correlation with their size, there is also a significant
direct correlation between this energy density and redshift. However, the size
also anticorrelates with redshift, so we calculate the partial correlations between
all these parameters.
The Pearson partial correlation coefficients in the correlations between $u_{\rm eq}$,
$P_{1.4}$, $D$, and 1+$z$ are given in Table~3.

\begin{table}[h]
\caption{The correlation of (log) equipartition energy density $u_{\rm eq}$
with $P_{1.4}$, or $D$, or 1+$z$,  when other parameters are held constant}
\begin{tabular}{lllll}
\hline
Correlation & $r_{XY}$ & $r_{XY/U}$ & ${\cal P}_{XY/U}$ \\
N=118     & & $r_{XY/V}$ & ${\cal P}_{XY/V}$ & $r_{XY/UV}$\\
& & & &   ${\cal P}_{XY/UV}$\\
\hline
$u_{\rm eq}-P_{1.4}/D      $ & +0.866 & +0.948 & $\ll$0.0001\\
$u_{\rm eq}-P_{1.4}/$1+$z$ & & +0.765 & $\ll$0.0001\\
$u_{\rm eq}-P_{1.4}/D$,1+$z$ & & & & +0.897\\
 & & & & $\ll$0.0001\\
$u_{\rm eq}-D/L_{1.4}$ & $-$0.802 & $-$0.925 & $\ll$0.0001\\
$u_{\rm eq}-D/$1+$z$    & & $-$0.830 & $\ll$0.0001\\
$u_{\rm eq}-D/L_{1.4}$,1+$z$ & & & & $-$0.925\\
& & & & $\ll$0.0001\\
$u_{\rm eq}-$(1+$z$)/$D$ & +0.631 & +0.690 & $\ll$0.0001\\
$u_{\rm eq}-$(1+$z$)/$P_{1.4}$  & & $-$0.059 & 0.55\\
$u_{\rm eq}-$(1+$z$)/$D,P_{1.4}$ & & & & +0.015\\
& & & & 0.86\\
\hline
\end{tabular}
\end{table}

\begin{figure*}[t]
\resizebox{180mm}{!}{\includegraphics{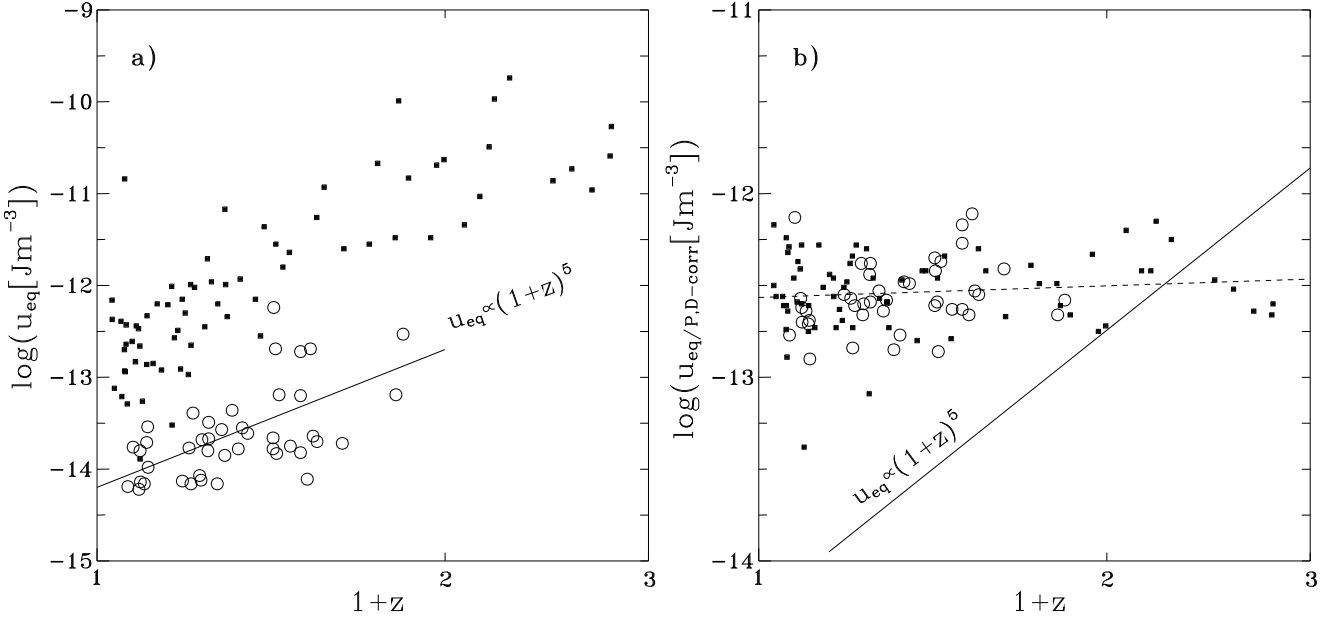}}
\caption{{\bf (a)} Equipartition energy density vs. redshift, {\bf (b)} the same energy
density transformed to the reference size of 400 kpc and 1.4-GHz total luminosity of
10$^{26}$ W\,Hz$^{-1}$. GRGs and NSGs are marked with the same symbols as in Fig.\,1.
The solid line in (a) and (b) indicates the presumed IGM pressure evolution
$p_{\rm IGM}\propto u_{\rm eq}\propto (1+z)^{5}$. The dashed line in (b) shows the
least squares fit to the transformed data}
\end{figure*}

The partial correlations coefficients in Table~3 clearly exhibit a strong
dependence of energy density (and so probably of average internal pressure) on
both the total radio luminosity and the source's size.  When these two parameters
are kept constant, the apparent correlation between $u_{\rm eq}$ and redshift
practically disappears. Some consequences of this effect are discussed in Sect.~4.

The direct correlation between $u_{\rm eq}$ and (1+$z$) in our sample is shown
in Fig.~4a. The solid line indicates the presumed IGM pressure evolution in the form
$p_{\rm IGM}\propto u_{\rm eq}\propto (1+z)^{5}$. Fitting a surface to the values
of log\,$u_{\rm eq}$ over the log$P_{1.4}$--log$D$ plane (where $P_{1.4}$ is in
W\,Hz$^{-1}$ and $D$ in kpc), we find

\begin{equation}
u_{\rm eq}\propto P^{0.65\pm 0.03}_{1.4}D^{-1.33\pm0.05}.
\end{equation}

\noindent
The above relation  does not differ much from that expected using the canonical
formula. However, the difference between the powers of $P$ and $D$ can be real
and justified by the non-constant axial-ratio parameter of the sources' cocoon,
$AR$, and by the fact that $P$ and $D$ are not independent variables. Indeed,
assuming that the values of both $AR$ and $D$ are
a function of the source age, $t$, and taking $AR\propto t^{0.23\pm 0.03}$
(Machalski et al. 2004) as well as $D\propto t^{3/(5-\beta)}$ with $\beta$=1.5
(cf. Kaiser et al. 1997; Machalski et al. 2004), one can find
$AR(D)\propto D^{0.27\pm 0.03}$. Because the cocoon volume is
$V\propto D^{3}AR^{-2}$, then $V^{-4/7}\propto D^{-1.4\pm 0.1}$. Also as the
luminosity of sources (according to all dynamical models) is time dependent, the
power of $P_{1.4}$ in Eq.\,(3) may differ from the value of 4/7 if the samples
comprise radio sources observed at different ages. 
  
Using Eq. (3), we transform $u_{\rm eq}$ values (these values for the
GRGs from our sample are given in Table~4 of Paper II, while those for the GRGs and
NSGs from the sample of Machalski et al. (2004) are recalculated for $H_{0}$=71 
km\,s$^{-1}$Mpc$^{-1}$ and $\Omega_{\rm m}$=0.27) to a reference 1.4-GHz luminosity
of 10$^{26}$ W\,Hz$^{-1}$ and size of 400 kpc. The relation between the transformed
energy density and redshift of the sample sources with the regression line on the
redshift axis is shown in Fig.~4b.

\subsubsection{Magnetic field partial correlations}

The Pearson partial correlation coefficients calculated for the correlations between
the equipartition magnetic field strength, $B_{\rm eq}$ and the total radio luminosity,
$P_{1.4}$, redshift, 1+$z$, and the source size, $D$, again confirm that the
strongest (anti)correlation occurs between $B_{\rm eq}$ and $D$ (logarithmic scales).
For N=118, the probability of a chance correlation is less than 0.0001.

\begin{figure}[t]
\resizebox{\hsize}{!}{\includegraphics{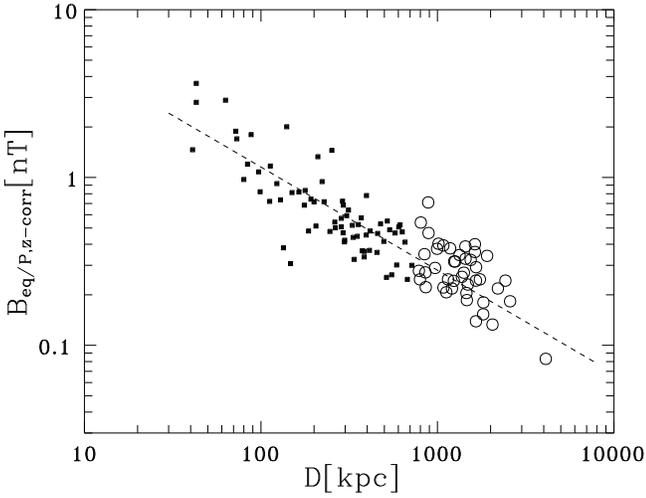}}
\caption{Equipartition magnetic field strength, transformed to the reference 1.4-GHz
luminosity of 10$^{26}$ W\,Hz$^{-1}$ and redshift of 0.5, vs. source size. GRGs and
NSGs are marked with the same symbols as in Fig.\,1. The dashed
line indicates the least-squares fit to the transformed data}
\end{figure}

As we did for the energy density parameter, a power-law dependence of
$B_{\rm eq}$ on $P_{1.4}$, and (1+$z$) values has been derived. Consequently
$B_{\rm eq}$ values, transformed to the reference 1.4-GHz luminosity of 10$^{26}$
W\,Hz$^{-1}$ and redshift of 0.5, are plotted against source size ($D$) in Fig.~5.
Though a dependence of the equipartition magnetic field on the source size
is expected, we show this plot because, according to the dynamical model of Kaiser
et al. (1997) and its application to observational data given in Machalski et al.
(2004), it reflects a more fundamental dependence of the lobes' (or cocoon) energy
density and the mean magnetic field strength on the dynamical age of radio sources.
We would like to emphasize a partial dependence of some observational
parameters of the sources, e.g. the total luminosity and size (referred here to as
fundamental parameters), on their age. Besides, these two parameters depend also
on the energy delivered to the lobes by the jets, as well as the density of the
ambient environment. Though we are not able yet to determine that age for the entire
sample of sources analysed in this paper, a subset of those sample sources with a
very similar linear size of about 300 kpc, and different ages and equipartition
magnetic fields, can be selected from Machalski et al. (2004). This subset is given
in Table~4, where all columns are self-explanatory; the size $D$ is recalculated
using the cosmological constants applied in this paper.
The entries in Table~4 clearly show the dependence of $B_{\rm eq}$ on the age,
when $D$ is held constant.

\begin{table}[h]
\caption{Example of the sample sources showing the correlation between $B_{\rm eq}$
and their age}
\begin{tabular}[80mm]{lccc}
\hline
Source  & $D$[kpc] & $t$[Myr] & $B_{\rm eq}$[nT]\\
\hline
3C437    & 316 & 6.4 & 4.38\\
3C322    & 283 & 7.3 & 3.28\\
3C267    & 315 & 12  & 2.93\\
3C244.1  & 294 & 14  & 1.72\\
3C337    & 297 & 24  & 2.20\\
3C357    & 296 & 27  & 0.60\\
3C319    & 297 & 43  & 0.71\\
0828+324 & 296 & 59  & 0.24\\
\hline
\end{tabular}
\end{table}

\subsection{Polarisation and depolarisation characteristics}

The rotation measure, $RM$, and depolarisation measure, $DP$, are closely related
to the distribution of thermal plasma and magnetic fields both inside and outside
the sources. The basic theory (Burn 1966; Gardner \& Whiteoak 1966) predicts that
a rotation of the polarisation plane without depolarisation would indicate a
foreground-resolved Faraday screen, whereas a
rotation accompanied by depolarisation would suggest a foreground screen as
well as a screen local to the sources. In the sample of Goodlet \& Kaiser (2005)
(which constitute  part of our NSGs sample) the authors found that both the
measured dispersion of $RM$ and the $DP$ correlate with redshift concluding that
their small-scale variations of $RM$ are caused by a local screen.

Our analysis, based on the limited polarisation data for the GRGs sample taken at
two frequencies only, does not allow convincingly constrain the location of
possible screens. However, the correlation and partial correlation tests can show
whether rotation and depolarisation properties of giant-sized and normal-sized
radio galaxies are similar or not.

\subsubsection{Rotation measure partial correlations, and relation between
rotation and depolarisation}

The determination of the rotation measure, $RM$, and depolarisation measure, $DP$,
for the sample GRGs was described in Paper II. The $RM$ and $DP$ values for
the NSGs have been available only for the sample members studied by Garrington et
al., and Goodlet et al. $RM$ values are not available for the sources taken from
Ishwara-Chandra et al., reducing our statistics to 44 sample sources with the
rotation measure determined. For the partial correlation calculations, we take
the average of the $RM$ and $DP$ values determined in the original papers for the
lobes of the sample sources. The Pearson correlation coefficients
and partial correlation coefficients for the correlation between $RM$ and the
sources' fundamental parameters show that the rotation measure is not correlated with
any of the fundamental parameters, i.e. the redshift, radio luminosity, and size.
Thus the $RM$ values for the GRGs would support the conclusion drawn
by Goodlet \& Kaiser (2005) that most of the observed $RM$ is caused by the Galactic
magnetic field and not by a Faraday screen local to the sample sources.

However all sample sources are also depolarised. In principle, the measured
depolarisation accompanied by a  rotation of the polarisation plane can tell
us something about
the matter and magnetic fields in the source itself and/or between it and the
observer. In practice, it is very difficult to recognize possible Faraday
screens acting for a given radio source (cf. Laing 1984). Regardless of possible
inner and/or outer screens, a decrease of depolarisation with $\lambda^{2}$ at
short wavelengths should accompany an increase of rotation with $\lambda^{2}$.

The plot of $DP$ vs. $\mid$$RM_{z}$$\mid$ (i.e. corrected to the sources' frame
by multiplying the measured values by (1+$z$)$^{2}$) for the lobes of the sources
with available $RM$ values,
shown in Fig.~6, indicates that the GRGs are, on average, less depolarised with
the polarisation plane less rotated than the corresponding characteristics of the
NSGs. This would suggest that the Faraday depth of intervening environment
surrounding GRGs (their lobes or cocoon) is lower in comparison to that around
less extended structure of NSGs.
As the rotation and depolarisation measures are probably related, we analyse
below how the $DP$ values in our sample are correlated with the principal
parameters, especially with the linear size.

\begin{figure}[t]
\resizebox{\hsize}{!}{\includegraphics{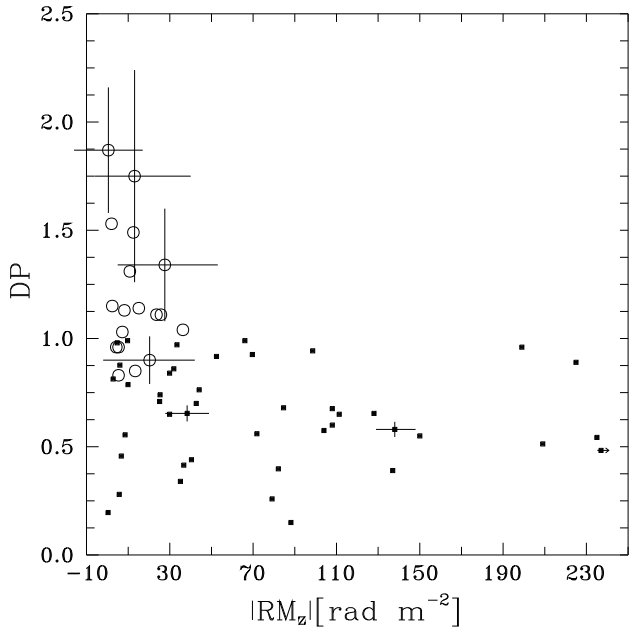}}
\caption{Depolarisation measure vs. absolute value of rotation measure corrected
for redshift, i.e. transformed to the sources' frame.  Crosses
show a typical error in both measures}
\end{figure}

\subsubsection{Depolarisation measure partial correlations}

If the measured depolarisation, $DP$, or a part of it was caused by a screen
local to the source, we would expect that $DP$ may correlate with $D$. This is the
case; the histograms of $DP$ values in three ranges of $D$ of the GRG and NSG radio
galaxies investigated are shown in Fig.~7. Note that, according to the adopted
definition of the depolarisation measure, an increase of the $DP$ values means a
decrease of the source's depolarisation. However, as the $DP$ values can also correlate
with the other fundamental parameters, we calculate the relevant Pearson
correlation and partial correlation coefficients, and there are given in Table~5.

\begin{table}[h]
\caption{The correlation of depolarisation measure $DP$ with (log) $D$, or $P_{1.4}$,
or 1+$z$,  when other parameters are held constant}
\begin{tabular}{lllll}
\hline
Correlation & $r_{XY}$ & $r_{XY/U}$ & ${\cal P}_{XY/U}$ \\
N=64     & & $r_{XY/V}$ & ${\cal P}_{XY/V}$ & $r_{XY/UV}$\\
& & & &   ${\cal P}_{XY/UV}$\\
\hline
$DP-D/P_{1.4}$  & +0.59 & +0.47 & $<$0.0001\\
$DP-D/$1+$z$ &  & +0.59 & $\ll$0.0001\\
$DP-D/P_{1.4}$,1+$z$ & & & & +0.47\\
 & & & & 0.0002\\
$DP-P_{1.4}/D$ & $-$0.41 & $-$0.01 & 0.97\\
$DP-P_{1.4}/$1+$z$    & & $-$0.45 & 0.0002\\
$DP-P_{1.4}/D$,1+$z$ & & & & $-$0.16\\
& & & & 0.61\\
$DP-$(1+$z$)/$P_{1.4}$ & $-$0.15 & +0.26 & 0.04\\
$DP-$(1+$z$)/$D$  & & +0.14 & 0.27\\
$DP-$(1+$z$)/$P_{1.4},D$ & & & & +0.21\\
& & & & 0.11\\
\hline
\end{tabular}
\end{table}

\begin{figure}[t]
\resizebox{\hsize}{!}{\includegraphics{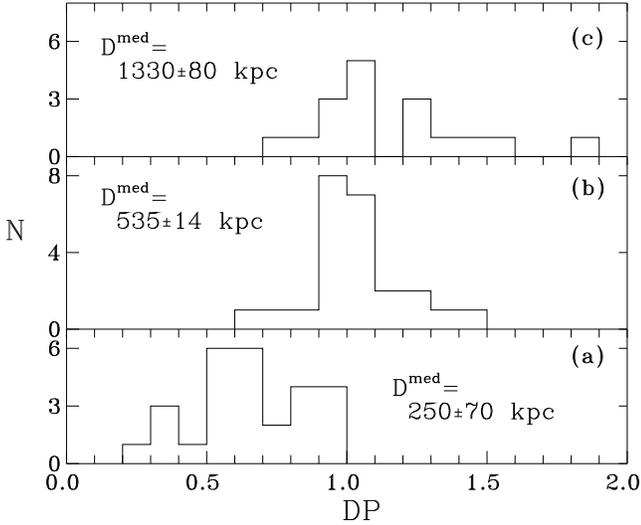}}
\caption{Histograms of the depolarisation measure $DP$ for {\bf (a)} radio galaxies
from the samples of Garrington et al. (1991) and Goodlet et al. (2004); {\bf (b)}
radio galaxies from the sample of Ishwara-Chandra et al. (1998); and {\bf (c)} giant
radio galaxies from our sample}
\end{figure}

The above tests confirm a significant correlation of $DP$ with $D$, and show a
residual $DP$--log(1+$z$) correlation. Fitting a surface to the $DP$ values over
the log$P_{1.4}$--log(1+$z$) plane, we find

\begin{equation}
DP\propto P_{1.4}^{-0.23\pm 0.06}(1+z)^{1.1\pm 0.6}.
\end{equation}

\noindent
In spite of the very uncertain dependence of $DP$ on 1+$z$ when $P_{1.4}$ and $D$
are held constant, we transform the
$DP$ values into the reference values of $P_{1.4}$=$10^{26.5}$ W\,Hz$^{-1}$ and
$z$=0.5. The $DP$ values corrected in this way are plotted against $D$ in Fig.~8.
As a result, larger radio galaxies tend to be less depolarised than smaller ones,
suggesting again that their depolarisation may be caused by a thin IGM local
to the sources. The statistical significance of this effect is very high (cf.
Table~5).

\begin{figure}[t]
\resizebox{\hsize}{!}{\includegraphics{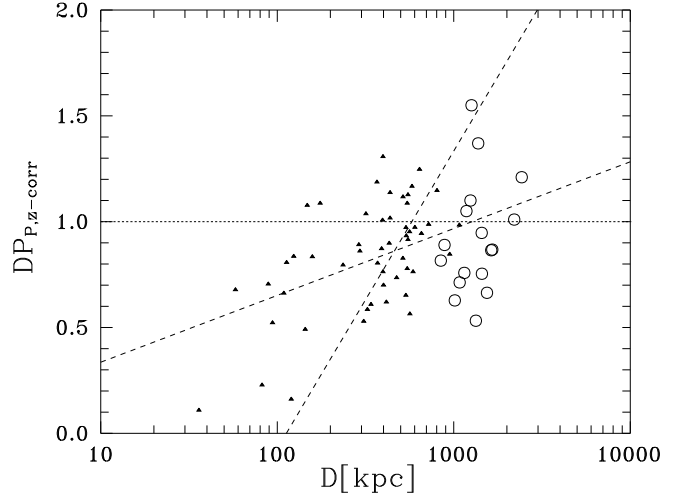}}
\caption{Depolarisation measure between 1.4 GHz and 4.86 GHz vs. linear size
of the sample sources. GRGs and NSGs are marked with the same symbols as in
Fig.~1. The dashed lines indicate the least-squares linear regression of the
data points on the abscissa and ordinate axes. The partial correlation
coefficient $r_{DP,D/P,z}$ is +0.47 (cf. Table~5)}
\end{figure}

\section{Discussion of the results and conclusions}

The important results of Section 3 are summarized in Table~6. In this Section,
we discuss some properties of the giant-size radio galaxies that have emerged
from our analysis.

\begin{table*}[ht]
\caption{Summary of the trends and correlations. The correlations are considered
significant (Yes) if the probability of the observed result under the null
hypothesis is $<$1\%, marginally significant (Yes?) if it is $<$5\%, and not
significant (No) otherwise}
\begin{tabular*}{178mm}{ll}
\hline
Dependence tested & Significant?\\
\hline
Core power is correlated with total radio luminosity & Yes(99.99\%)\\
Core power is correlated with redshift         & No\\
Core power in GRGs is higher than that in NSGs & No\\
Core prominence is correlated with total radio luminosity & No\\
Core prominence is correlated with redshift         & No\\
Core prominence is correlated with surface brightness of the lobes or cocoon  & Yes(99.9\%)\\
Energy density is correlated with total radio luminosity  & Yes(99.95\%)\\
Energy density is correlated with redshift          & No\\
Energy density is correlated with linear size       & Yes(99.99\%)\\
Energy density in GRGs is lower than that in NSGs   & Yes(99.9\%)\\
Equipartition magnetic field is correlated with linear size & Yes(99.99\%)\\
Fractional polarisation in GRGs is lower than that in NSGs  & No\\
Rotation measure in GRGs is lower than that in NSGs & ?\\
Rotation measure is correlated with fundamental parameters (total radio
luminosity, linear size, redshift)  & No\\
Rotation measure is correlated with depolarisation  & Yes?\\
Depolarisation measure is correlated with linear size & Yes(99.98\%)\\
Depolarisation measure is correlated with redshift    & No?(89\%)\\
\hline
\end{tabular*}
\end{table*}

\subsection{Core power and core prominence}

The core power is highly correlated with the total radio luminosity of FRII-type
radio sources, even if the influence of other fundamental parameters (the linear
size and redshift) on the above correlation is eliminated. The core powers of
GRGs do not differ from those of NSGs. However, the core prominence parameter
does not depend on the total power, but anti-correlates with  energy density
in the lobes or cocoon of the sample sources. On the other hand, the energy density
ought to evolve with the source age (cf. the dynamical models of Kaiser \& Alexander
1997; Blundell et al. 1999; Manolakou \& Kirk 2002).
This implies that a dynamical age of the radio structure is a more fundamental
parameter than its radio luminosity and size.

\subsection{Energy density, internal pressure, and their implication for the
hypothesis of the IGM pressure evolution with redshift}

The former studies (e.g. Arnaud et al. 1984; Rawlings 1990) indicated that the
minimum internal pressures in diffuse lobes and bridges of FRII-type radio galaxies
equal the pressure of IGM in cases where detectable X-ray emitting gas surrounds
the radio structure. Moreover, the studies also showed that the diffuse radio structures
located outside these high-density environments may be in thermal equilibrium with
the ambient medium whose emissivity cannot be directly determined. Therefore, the
approximate equality of the derived internal and external pressures justifies the
energy equipartition assumption that was, and is usually used in calculation of
internal pressure within the radio lobes.

The expected electron pressure in the adiabatically expanding Universe is
$p_{\rm IGM}=p^{0}_{\rm IGM}(1+z)^{5}$ with  $p^{0}_{\rm IGM}=2\cdot 10^{-15}$
N\,m$^{-2}$ (cf. Subrahmanyan \& Saripalli 1993). On the other hand, analytical
models of the dynamical evolution of FRII-type sources (e.g. Kaiser \& Alexander
1997) assume that their internal pressure depends on the source's size, hence is
a function of its age (cf. eq.\,(2) in Kaiser 2000).

The statistical test in Sect.~3.3.1 shows that $u_{\rm eq}$ (thus likely the
cocoon internal pressure) is independent of redshift when the radio luminosity
and size are kept constant. If the tenuous material in the cocoon of GRGs attains
an equilibrium state and its pressure equals the pressure of the IGM, the above
result will disagree with the expected cosmological evolution of the IGM.
Another possibility is that the cocoon, even in the largest sources, is still
overpressed with respect to the surrounding medium. Therefore our result is
identical with that of Schoenmakers et al. (2000), who concluded that there was
no evidence in their sample for a cosmological evolution of  energy density
in the lobes of GRGs, and there was therefore also no evidence for a
cosmological evolution of pressure within the IGM. We also agree with their
conclusion that a rejection of the hypothesis of the IGM pressure evolution
proportional to (1+$z$)$^{5}$ would be possible if high-redshift GRGs (at
$0.6<z<1$) with energy densities less than about $2\times 10^{-15}$ N\,m$^{-3}$
were discovered.

\subsection{Polarisation}

Global polarisation characteristics of the sample GRGs are similar to those of NSGs.
The only trends (however of low statistical significance due to the low number of
sources in the samples used) are:

\noindent
-- the dispersion of the rotation measure of GRGs is lower than that of NSGs, and

\noindent
-- GRGs tend to be less depolarised than NSGs.

\noindent
Thus, taking also into account the significant correlation of the depolarisation measure
with the source linear extent, all these characteristics suggest that a part of the
rotation and depolarisation is caused by a Faraday screen local to the extragalactic
FRII-type radio sources.

Because the low depolarisation and rotation measures determined for GRGs describe the
polarised emission from their lobes, the above implies that the IGM surrounding the
lobes (or cocoon) of GRGs is evidently less dense than that in a vicinity of NSGs.
Obviously, these global characteristics, determined at two observing frequencies
only, tell us nothing about the geometry and composition of the intervening material.
Further analysis of polarisation asymmetries between the lobes can be more
promising, which we intend to perform in a separate paper.

\begin{acknowledgements}

We are grateful to Dr Peter Barthel for constructive comments which helped us to
improve the paper and clarify our results. 
This work was supported in part by the State with funding for scientific research
in years 2005-2007 under contract No. 0425/PO3/2005/29
 
\end{acknowledgements}

\end{document}